\documentclass[referee]{cjaa}           
\usepackage{graphicx}                   
\input{epsf.sty}                        
\input{psfig.sty}                       

\begin{document}

   \title{Implementation of new OPAL tables in Eggleton's stellar
evolution code
}

   \volnopage{Vol.0 (200x) No.0, 000--000}      
   \setcounter{page}{1}          
   \author{Xuefei Chen
      \inst{1}\mailto{}
   \and Christopher A. Tout
      \inst{2,3}
      }
   \offprints{Xuefei Chen}                   
   \institute{National Astronomical Observatories/Yunnan Observatory, CAS, Kunming, 650011, P.R.China\\
             \email{xuefeichen717@hotmail.com}
        \and
             Institute of Astronomy, The Observatories, Madingley Road,
Cambridge CB3 0HA, England
\and
Centre for Stellar and Planetary Astrophysics, School of Mathematics,
Monash University, Clayton, Victoria 3800, Australia
\\
          }

   \date{Received~~2006 month day; accepted~~2006~~month day}

   \abstract{ Based on the work of OPAL (Iglesias \& Rogers 1996) and
   Alexander \& Ferguson (1994), we construct a series of opacity
   tables for various metallicities $Z=0$, $0.000\,01$, $0.000\,03$,
   $0.000\,1$, $0.000\,3$, 0.001, 0.004, 0.01, 0.02, 0.03, 0.04, 0.05,
   0.06, 0.08 and 0.1.  These tables can be easily used in Eggleton's
   stellar evolution code in place of the old tables without changing
   the code.  The OPAL tables are used for ${\log}_{10}(T/{\rm K})>
   3.95$ and Alexander's for ${\log}_{10}(T/{\rm K})< 3.95$.  At ${\rm
   log}_{10}(T/{\rm K})=3.95$, the two groups' data fit well for all
   hydrogen mass fractions.  Conductive opacities are included by
   reciprocal addition according to the formulae of Yakovlev \& Urpin
   (1980).  For different metallicities $1\,M_\odot$ models are
   constructed with the new opacity tables.  Comparison of $1$~and
   $5\,M\odot$ models constructed with the older OPAL tables (Iglesias
   \& Rogers 1992) shows that the new opacities have most effect in
   the late stages of evolution, the extension of the blue loop during
   helium burning for intermediate-mass and massive stars.
 }

   \authorrunning{Chen and Tout }            
   \titlerunning{New Opacity in PPE code }  

   \maketitle

\section{Introduction}           
\label{sect:intro}
In the 1970's Peter P. Eggleton (\cite{egg71,egg72,egg73}) devised a
stellar evolution code (PPE code in this paper), which has been
updated with the latest physics over the last three decades (Han et
al.~\cite{han94}; Pols et al.~\cite{pol95}; Pols et al.~\cite{pol98}).
The PPE code is now used worldwide because of its convenience,
expediency and easy application to binary evolution.
For example, this code and its stripped-down version EZ
(rewritten by Bill Paxton in Fortran 90) are the main stellar evolutionary
codes of MODEST (Modeling Denser Stellar system),
which is a loosely knit collaboration between various groups 
working in stellar dynamics, stellar evolution and stellar hydrodynamics. 

Opacity is a measure of the degree to which matter absorbs photons, and
it significantly affects stellar evolution. Opacity tables are therefore
important for a stellar evolution code.  Four main processes influence
the opacity in a star: the transition of an electron between two
energy levels in an atom, ion or molecule (bound--bound), ionization
of an atom or ion by an incoming photon with enough energy
(bound--free), an electron and photon interacting near an atom or ion
(free--free) and electron scattering.  
Composition, temperature and density all affect the opacity, 
and the calculation of opacities in a star is
complicated and difficult.
With the models of the above process,
the OPAL group presented their original opacity tables in 1992
(Iglesias \& Rogers \cite{ir92}) (hereinafter OPAL92). 
Since then they have further considered the details of their
models (Iglesias \& Rogers~\cite{ir96}, IR96) 
and presented new opacity tables\footnote{available
at http://www-phys.llnl.gov/Research/OPAL} (hereinafter OPAL96).
In comparison with OPAL92, OPAL96 has been improved in the physics 
and numerical procedures as well as some corrections to their code.
Here we just give a brief description on the improvement in the physics, 
one may get more details in the paper of IR96. 
The main feature of OPAL96 is the inclusion of seven more metals
in addition to the original 12 metals.
Full intermediate coupling is used for Ar and heavier elements while 
LS coupling is used for the lighter elements.
The absorption and scattering from various H and He species are included
(only the formation of ${\rm H_2}$ is considered in OPAL92).
Also the red wings of line profiles are included and the effect of 
neutral  H and He on spectral line broadening are considered.
For ionization balance, the Debye-H\"uckle approximation is used in OPAL92
while higher order Coulomb terms are included in OPAL96.
All these modifications given above will affect the opacities.
The inclusion of 19 metals leads to an increase in opacity 
by as much as 20 per cent
for population I with solar metallicity $Z=0.02$.
The opacity changes induced by other modifications are less than 10 per cent.

The opacity changes will affect the stellar structure and evolution,
especially for intermediate-mass and massive stars,
as has been discussed by several previous papers
(Alongi et al. \cite{al93}, Stothers \& Chin \cite{sc94}).
The models with OPAL96 will also exhibit some differences 
in structure and evolution compared to those with OPAL92.
These differences are probably not very large (see Sect.3 in this paper).
However, OPAL96 has become the most comprehensive opacities 
available today.
In IAUS 241, which will be held from 10--16 December 2006 in Spain,
there are two so-called challenges, and one is for stellar evolution.
In the stellar evolution challenge, 
OPAL96 and the data from Alexander \& Ferguson ~(\cite{af94}) 
are appointed as opacity inputs in physical specifications.  
Meanwhile, as a stellar evolutionary code, 
it is necessary to improve it with new physics and new physical inputs,
and researchers are always willing to 
apply these in order to
to improve the accuracy of evolutionary calculations.

Eldridge \& Tout~(\cite{et04}) 
took a detailed look at, and implemented the opacity for enriched carbon
and oxygen mixtures with OPAL96.
They also provided a new stellar code with a new routine 
to read their opacity tables.
In particular they use the variable $R={\rho}/T_6^3$, where  $\rho$
is the density, $T_6=T/10^6\,K$ and $T$ is the temperature.
Then ${\log}_{10}(T/{\rm K})$ ranges from 3.0 to 10 
and ${\log}_{10}(R/{\rm g\,cm^{-3}})$ from $-8.0$ to 7.0.
At low temperatures (${\log}_{10}(T/{\rm K})< 4.0$), 
they use the data of Alexander \& Ferguson~(\cite{af94}) 
and at high temperatures ($8.7 < {\log}_{10}(T/{\rm K}) < 10$) they
fill
the table using the fits of Iben (\cite{ib75}).
The major conclusion of their study is that 
the changes induced 
by properly including opacities for varying C and O mixtures are small,
but may improve the numerical stability in the late stages of evolution, 
because they resolve changes with composition in greater detail.
Their code and tables are available at 
{\it http://www.ast.cam.ac.uk/stars}.

At present, many astronomers are still using the old version of the code
and the most widely used opacity tables in it were constructed by Pols et al.
(\cite{pol95}) from OPAL92.
The metallicities, $Z$, for these tables are
0.0001, 0.0003, 0.001, 0.004, 0.01, 0.02 and 0.03.
These Z values may fulfill the values necessary for most studies.
However, for some special objects, 
i.e. the objects of Population III or those in the center of the Milky Way
(about 3 factors of solar metallicity),
it is necessary to enlarge the range of Z values.
Therefore we have constructed a series of opacity tables 
in the form of Pols et al. (\cite{pol95}), 
in (${\log\, \rho}$, $\log\,  T$) space from OPAL96.
The $Z$ values
specifically calculated are 
0, $0.000\,01$, $0.000\,03$, $0.000\,1$, $0.000\,3$,
0.001, 0.004, 0.01, 0.02, 0.03, 0.04, 0.05, 0.06, 0.08 and 0.1.
These tables can be implemented in the PPE code 
just as the old opacity tables, and
have no influences on stellar evolution 
in comparison to those from more detailed C and O mixtures
except for numerical stability in the late stages as shown in 
Eldridge \& Tout~(\cite{et04}). 
We present
zero-age main-sequence models of
various metallicities, $Z$, and some evolved models
for examination and comparison to models made with other tables.
  
\section{Table construction}
\label{sect:Comp}
In OPAL96, 
${\log_{10}}(T/{\rm K})$ ranges from 3.75 to 8.70 
and ${{\log}_{10}}(R/\rm g\,cm^{-3})$ 
from $-8$ to 1.
Alexander \& Ferguson~(\cite{af94}) derived molecular \& grain opacities
for low temperatures 
(${\log_{10}}(T/{\rm K})$  from 3.0 to 4.1 ).
We use OPAL tables for ${\log_{10}} (T/{\rm K}) > 3.95$ 
and Alexander's for ${\log_{10}}(T/{\rm K}) < 3.95$. 
At ${\log_{10}} (T/{\rm K}) = 3.95$, the two sets of data match well.
The radiative opacities, $\kappa_{\rm rad}$ 
(Rosseland mean opacity obtained from OPAL and Alexander tables), 
are supplemented with conductive opacities, $\kappa_{\rm con}$,
according to the formulae of Yakovlev \& Urpin (\cite{yu80}).
The opacity $\kappa$ is obtained by

\begin {equation}
{1\over\kappa} = {1\over\kappa_{\rm rad}} + {1\over\kappa_{\rm con}}.
\end{equation}
 
We construct a rectangular table
in ${\log\, }\rho$, ${\log\, }T$ space with
$-12 < {\log}_{10}(\rho/{\rm g\,cm^{-3}}) < 10$ in steps of 0.25 and 
$3 < {\log}_{10}(T/{\rm K}) < 9.3 $ in steps of 0.05.
At very high temperature (${\log_{10}}(T/{\rm K}) > 8.7$),
the opacity is dominated by electron scattering, and
relativistic effects are important. 
We fill this part with the formula of Iben (\cite{ib75}),
\begin {equation}
\kappa_{\rm e} = [0.2-D-(D^2+0.004)^{1/2}](1+X),
\end{equation}
where $D={\log}_{10}T_6-1.7$.
 
Outside the ranges of $R$, values are obtained by extrapolation in the
old opacity tables.  These parts have no real validity for stellar
evolution, so we simply fill them with boundary values.  The $Z$ values
specifically calculated are 0, $0.000\,01$, $0.000\,03$, $0.000\,1$, $0.000\,3$,
0.001, 0.004, 0.01, 0.02, 0.03, 0.04, 0.05, 0.06, 0.08 and 0.1.  For
each $Z$, $(X,X_{\rm C}+X_{\rm O}) = (0.8, 0.0)$, (0.7, 0.0), (0.5, 0.0),
(0.35. 0.0), (0.2, 0.0), (0.1, 0.0), (0.0, 0.0), (0.0, 0.5), (0.0,
0.8) and (0.0, $1-Z$), where $X$ is the hydrogen mass fraction and $X_{\rm
C}+X_{\rm O}$ is the sum of enhanced carbon and oxygen following helium
burning.

\section{Results and Comparisons}
\label{sect:data}
\begin{figure}
\centerline{\psfig{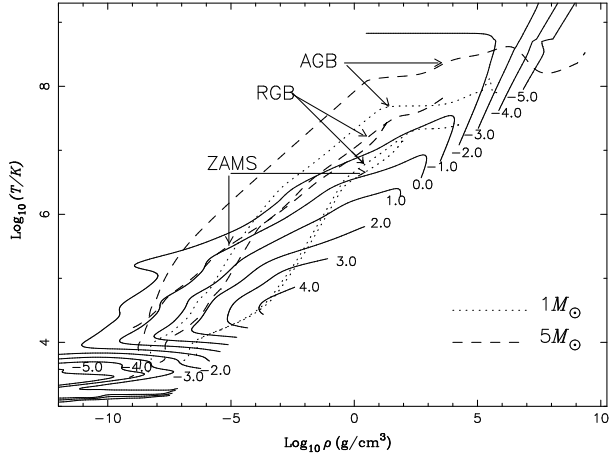}}
\caption{Contours of ${\log\, \kappa}$ 
in (${\log\, }T$,${\log\, }\rho$) space at
${\log}_{10}({\kappa}/{\rm cm^2\,g^{-1})} = 4.0$,
3.0, 2.0, 1.0, 0.0, $-1.0$, $-2.0$, $-3.0$, $-4.0$ and  $-5.0$.
The dotted lines show the internal structure of a $1M_{\odot}$ star 
in this space at different evolutionary
stages: zero-age main sequence (ZAMS), red giant branch (RGB) and 
asymptotic giant branch (AGB).
The dashed lines are similar models of a $5M_{\odot}$ star.}
\label{c1}
\end{figure}

For $Z=0.02$ contours of ${\log\, }\kappa$ 
in (${\log\, }T,\,{\log\, }{\rho}$) space 
are presented in Fig.~\ref{c1}.
The structures of stars of $1M_{\odot}$ and $5M_{\odot}$ at different 
evolutionary stages: zero-age main sequence (ZAMS), red giant branch (RGB)
and asymptotic giant branch (AGB), are also shown in this figure.
We see that
the two stars stay in the ranges given by the OPAL and Alexander tables
throughout their lives.
For more massive stars, the density decreases and the temperature increases, 
so that the three lines move up to the left, far away from the boundary.

\begin{figure}
\centerline{\psfig{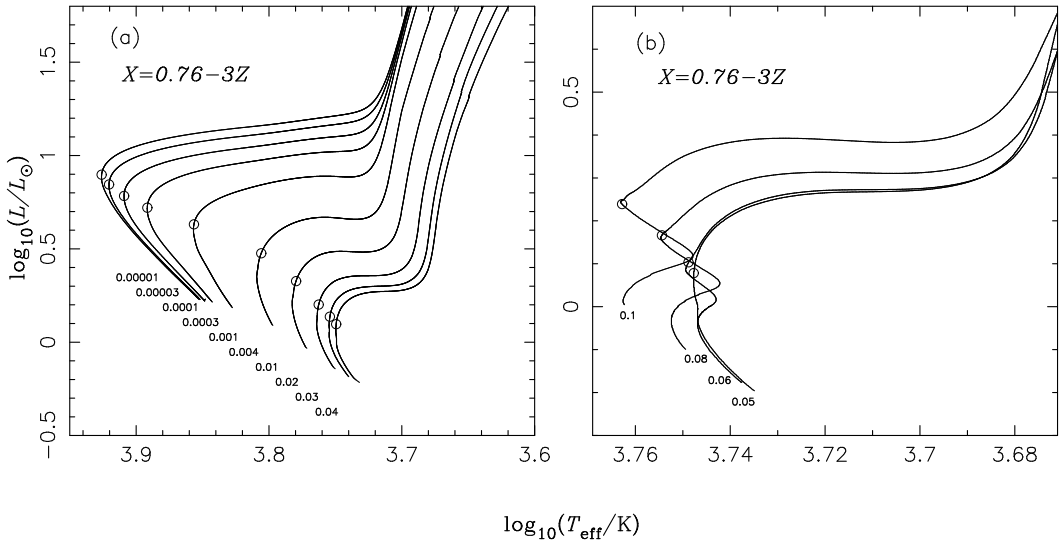}}
\caption{Evolutionary tracks for a star of $1\,M_\odot$ with different metallicities
($Z=0.000\,01$, $0.000\,03$, $0.000\,1$, $0.000\,3$, 0.001, 0.004, 0.01, 0.02, 
0.03, 0.04, 0.05, 0.06, 0.08 and 0.1).
The initial hydrogen mass fraction $X$ is $0.76-3Z$ in each case.}
\label{e1}
\end{figure}

\begin{figure}
\centerline{\psfig{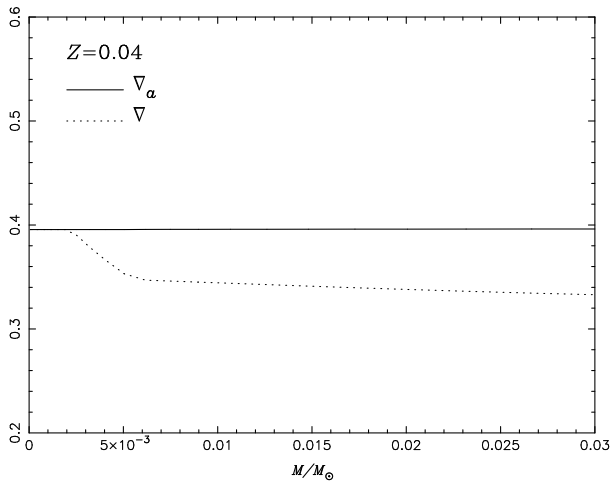}}
\caption{Profiles of adiabatic temperature gradient $\nabla_{\rm ad}$ 
and actual temperature gradient $\nabla$ in a star of 1$M_{\odot}$ 
with $Z=0.04$. }
\label{grad}
\end{figure}

Figure~\ref{e1} presents the evolutionary tracks for stars of $1\,M_\odot$
but different $Z$. The circles in the tracks are at the point when
$X\approx 0.001$ in the centre of the star.
The lifetime $t_{\rm MS}$, ${\log}_{10}(R/R_{\odot})$, 
${\log}_{10}(L/L_{\odot})$
and ${\log}_{10}(T_{\rm eff}/{\rm K})$ at the circles are listed in Table \ref{1}.
We can see core contraction 
when the central hydrogen is nearly exhausted for stars with $Z > 0.05$.
This indicates that a convective region develops in the centre
of these stars.
In fact a central convective region already exists at the end of 
main-sequence evolution 
in a star with $Z = 0.04$ (see Fig.~\ref{grad}).

Combining Table~\ref{1} and Fig.~\ref{e1}
we see that  $t_{\rm MS}$ decreases with luminosity 
as $Z$ ranges from $0.000\,01$ to $0.000\,3$. 
This differs from the usual expectations in stellar evolution -- 
larger luminosity results in a shorter lifetime on the main sequence.
This discrepancy is driven by the variation of opacity with $Z$ which
increases the size of the burning core even though
there is little difference in the central hydrogen mass fraction 
over this range of $Z$.

\begin{figure}
\centerline{\psfig{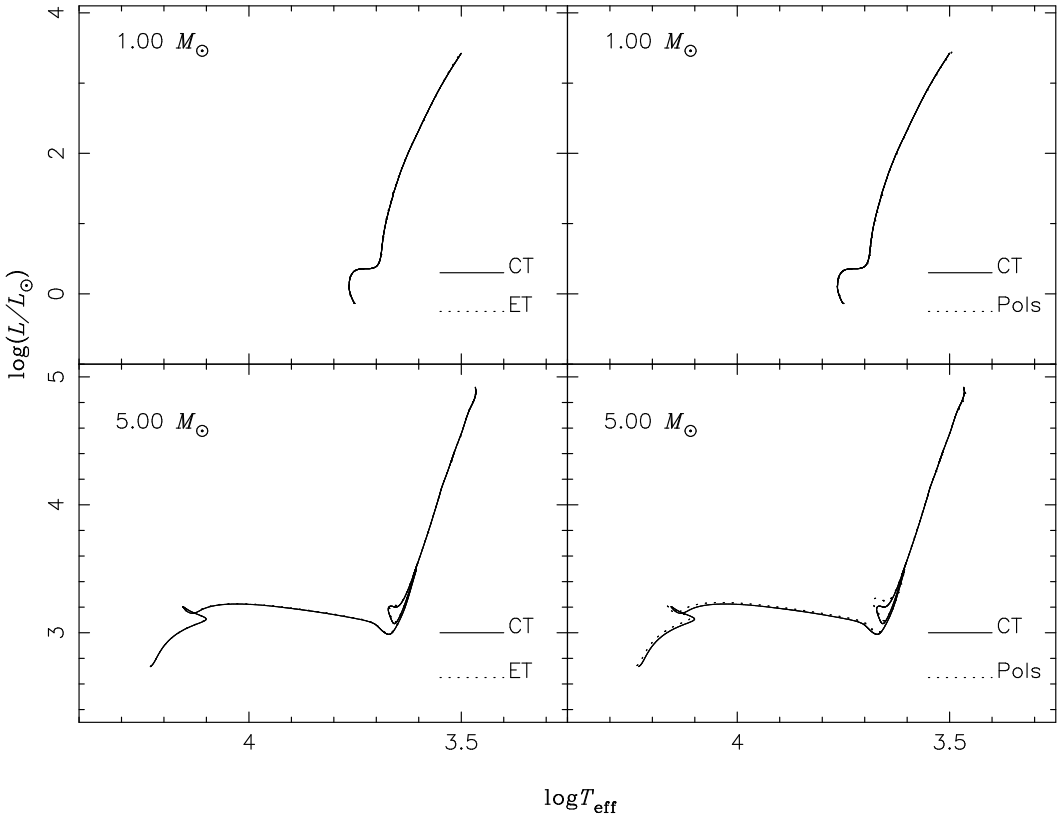}}
\caption{Evolutionary tracks of stars of $1M_{\odot}$ and $5M_{\odot}$ 
for different opacities for $Z=0.02$. The opacities are from 
Pols et al. (\cite{pol95}), Eldridge \& Tout (\cite{et04}) and this
work (CT).
For $1M_{\odot}$, the three are exactly overlapped.
For $5M_{\odot}$, the lines from ET and CT are also well matched.
}
\label{e2}
\end{figure}

\begin{table*}
\caption{Some parameters for the stars in Fig.\ref{e1} 
when the central hydrogen mass fraction is about 0.001.
The main-sequence lifetime $t_{\rm MS}$ is 
the time from zero-age main sequence to this point which is marked by
a circle on each track.}
\begin{tabular}{ccccccccccccccc}
\hline
 $Z$& $10^{-5}$& $3\times 10^{-5}$& $10^{-4}$& $3\times 10^{-4}$&
$10^{-3}$& $4\times 10^{-3}$& 0.01& 0.02& 0.03& 0.04& 0.05& 0.06&
0.08& 0.1\\ \hline
 $t_{\rm MS}/10^9\,\rm yr$ & 5.544 & 5.432& 5.324& 5.229& 5.310& 6.876& 7.560& 9.085
& 9.576 & 9.297 & 8.550& 7.922& 6.343& 4.500\\
 ${\log_{10}}R/R_{\odot}$ &  0.120&  0.105 &  0.097 &  0.100 &  0.126 &  0.150 
&  0.128  &0.099 &   0.083 & 0.073  &0.067 &0.078  &0.098 &0.117 \\
${\log_{10}}L/L_{\odot}$ &0.897 & 0.845 &  0.783 & 0.720 & 0.631 & 0.476 & 0.327
& 0.201 & 0.136 &  0.096 & 0.078 & 0.104 & 0.166 &  0.239\\
 ${\log_{10}}T_{\rm eff}/\rm K$  &3.926& 3.920 & 3.909 & 3.892 & 3.856 & 3.806 & 3.780 
& 3.763 &  3.754 & 3.749 & 3.748 & 3.749 & 3.754 &  3.763\\ 
\hline 
\label{1}
\end{tabular}
\end{table*}
        
For comparison, we show the evolutionary tracks of a star 
with $1M_{\odot}$ and $5M_{\odot}$ for different opacities 
when $Z=0.02$ in Fig.4. 
The solid lines give the results of this work (CT in the figure)
and the dotted lines that of Pols et al. (\cite{pol95}) (Pols)
and Eldridge \& Tout (\cite{et04}) (ET).
We see that 
the tracks for $1M_{\odot}$ precisely overlap. 
For $5M_{\odot}$, the tracks of CT and ET are also well-matched while
those of CT and Pols have some discrepancies on the main sequence and 
blue loop.
The divergences of CT and Pols are due to the 20 per cent
increase in opacity of OPAL96 in comparison with OPAL92.
As discussed by  Stothers \& Chin (\cite{sc94}),
the development and extension of blue loop are strongly dependent in opacity.
Since the main opacity changes are due to the inclusion of more metal
elements and the treatments on them,
the discrepancies of the results between OPAL92 and OPAL96 
for intermediate-mass and massive stars may be larger with increasing $Z$.
Therefore we can expect the new opacity tables to improve the accuracy of
stellar models of intermediate-mass and massive stars.

\begin{figure}
\centerline{\psfig{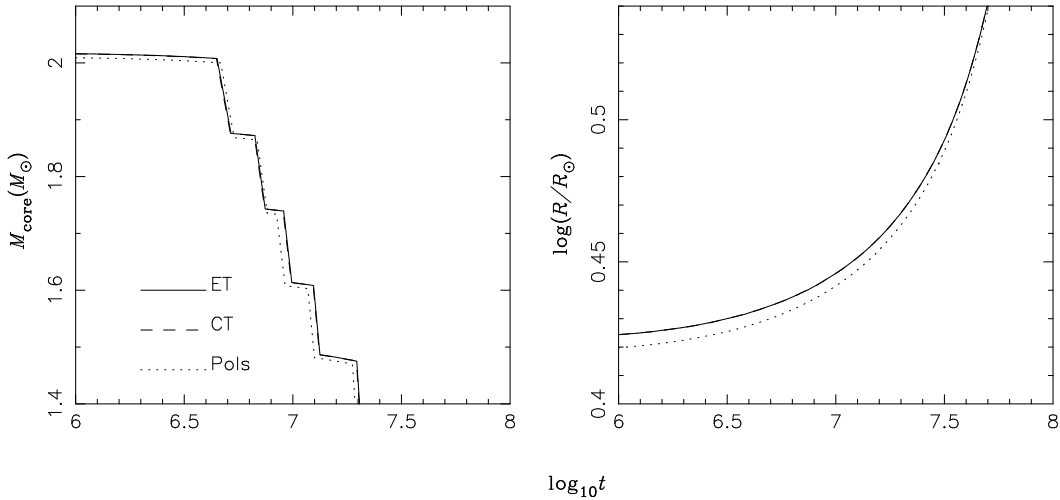}}
\caption{Core mass and stellar radius change with age
for the $5M_{\odot}$ model with different OPAL tables. 
The lines from ET and CT are precisely overlapped.}
\label{s1}
\end{figure}

The opacity changes may also affect stellar structures. 
The increase of opacity means an enhancement of energy fraction absorbed 
as the energy translates through a certain matter.
Therefore a star with higher opacities (i.e from OPAL96) 
has a lower luminosity and a larger radius 
as well as a smaller surface effective temperature. 
Another relatively important feature with opacity increase is 
the size of the central convective region 
for intermediate-mass and massive stars. 
The radiation temperature gradient $\nabla _{\rm rad}$ 
is proportional to opacity for certain conditions. 
Therefore the opacity increase will cause an increase of $\nabla _{\rm rad}$, 
as the convective criteria 
(whether for Ledoux criterion or Schwarzschild criterion) 
are fulfilled more easily. 
The enlargement of the central convective region 
will contribute to energy production, 
and thus to the surface luminosity of a star. 
A new equilibrium will be first built in the star with new opacities 
and the surface luminosity of the star will be still less than 
that with lower opacities (Fig.\ref{e2}) in the equilibrium.
The core mass and stellar radius for $5M_{\odot}$  model with OPAL92 and
OPAL96 are illustrated in Fig.\ref{s1}.
The models with OPAL96 (i. e ET and CT) 
have slightly longer timescales on the main sequence.
The age differences are about $5 \times 10^7$ yr and $3 \times 10^6$ yr for 
$1M_{\odot}$ and $5M_{\odot}$, respectively.
Alongi et al. (\cite{al93}) presented the influences of 
various opacity calculations on the stellar core mass 
and on the surface luminosity. 
Though the differences between OPAL92 and OPAL96 are not so obvious as
those between LAOL and OPAL92 as compared by Alongi et al. (\cite{al93}),
our discussions here are similar to those results.

Our tables and ZAMS models ($0.1 < M/M_\odot < 160$ for various metallicities) are available at {\it http://www.ynao.ac.cn/~bps}.
One may also send a request  for them to 
{\it xuefeichen717@hotmail.com} for them.

\section{Acknowledgments}
\label{sect:analysis}
The authors thank Dr O. R. Pols for his kind help.
The authors also thank the anonymous referee for his good suggestions
to improve the paper. 
Chen thank Pokorny R. S. for his improvement in language.
This work was supported
by the Yunnan Natural Science Foundation (Grant No. 2004A0022Q),
the Chinese Academy of Science Foundation (Grant No. O6YQ011001), 
the Chinese National Science Foundation (Grant NO.10433030) and 
the Chinese Academy of Sciences (No. KJCX2-SW-T06).

\label{lastpage}

\end{document}